\documentstyle[aps,psfig,twocolumn]{revtex} 
\input{epsf}

\newcommand{\bq}{\begin{equation}}
\newcommand{\ee}{\end{equation}}
\newcommand{\bea}{\begin{eqnarray}}
\newcommand{\eea}{\end{eqnarray}}

\begin{document}
\draft
\title{ Quantum wells, wires and dots with finite barrier:\\
 analytical expressions of the bound states 
}
\author{X. Leyronas$^{1}$, M. Combescot$^{2}$}
\address{ 
$1$ Laboratoire de Physique Statistique, Ecole Normale Sup\'erieure*, 
24 rue Lhomond, 75231 Paris Cedex 05, France\\ 
$2$ GPS, Universit\'e Denis Diderot et Universit\'e Pierre et Marie Curie,
CNRS,\\
 Tour 23, 2 place Jussieu, 75251 Paris Cedex 05 France.\\ 
}
\date{\today}
\maketitle
\begin{abstract}
From a careful study of the transcendental equations verified by
the bound states energies of a free particle in a quantum well, cylindrical wire or spherical dot with finite potential barrier, we have derived analytical 
expressions of these energies which reproduce impressively well the numerical 
solutions of the corresponding transcendental equations for {\it all}
confinement sizes and potential barriers, without any adjustable parameter. 
These expressions depend on a unique dimensionless parameter which 
contains the barrier height and the sphere, wire or well radius.
\end{abstract}
\pacs{PACS numbers: 73.21.Fg}

The present experimental and theoretical efforts in semiconductor physics
are essentially devoted to confined structures: quantum wells, wires or dots.
In most calculations, the potential barrier between the confined structure 
and the outside semiconductor is assumed to be infinitly high for simplicity.
When the barrier is finite, the number of bound states, which is infinite
for infinite barrier, can be reduced to $1$ or even $0$ for spherical dots,
each energy level being substantially modified by the usual experimental 
confinements.

The great advantage of the infinite barrier assumption lies in the fact that
the energies and wave functions have analytical expressions so that all
theoretical calculations using them are rather easy. The introduction of the
finite barrier height \cite{bastard,haug,banyai,rosencher,nag,zheng,fang,campi} makes everything much more complicated.
Even in the simplest case of one confined direction only, namely the quantum
well, the bound states energies are given through a transcendental equation.
Its solutions can be obtained either by a numerical calculation
or by "reading"  their values on a set of curves which makes their use
quite unconvenient in quantitative calculations.   
The cylindrical wire and spherical dot confinements being obviously more
complicated, the situation is even worse for the energy of these 2D and
3D geometries.

The purpose of this communication is to provide \emph{analytical expressions}
for the energies of \emph{any} bound state level of these three geometries,
\emph{valid for all confinements}\,, i.e. all barrier heights and 
confinement sizes. We will show that these energies have the same analytical
form, which depends on a unique dimensionless confinement parameter $\nu$,
in which enter the barrier height and the confinement extension.
Let us add that, once the energy levels are known, it is straightforward
to get from them the exact wave functions of all these bound states.

\textbf{\underline{1. Results:}}

We consider a particle of mass $m$ confined in a sphere or cylinder of 
radius $R$ or in a quantum well of width $2R$, the energy barrier being $V$. 
From $R$ and $V$, we can construct the dimensionless parameter $\nu$ which rules 
all the physics of these finite barrier problems, namely
\bq\label{eq1}
V=\hbar^2 \nu^2/2mR^2
\ee
The infinite barrier limit corresponds to $\nu$ infinite.
From this parameter $\nu$, we can already note that, as the physical scale for 
the barrier height is $\hbar^2 /2mR^2$, a given barrier $V$ between two 
semiconductors can appear as high or low depending on the confinement
extension: the larger the $R$, the better the $V=\infty$ approximation
for the same $V$. In other words, finite barriers effects are going to be very 
important for strongly confined systems.

Following Eq.(\ref{eq1}), we are led to measure the particle energies in the 
same unit as $V$, namely
\bq\label{eq2}
E=\hbar^2 \alpha^2/2mR^2=V-\hbar^2 \beta^2/2mR^2
\ee 
Note that for $\beta\simeq 0$, the energy is close to the top
of the well while for $\beta\simeq\nu$, $\alpha$ is close to $0$ so that the
level is deep inside the well.

From the transcendental equation verified by the various energy levels
of these confined geometries given below, we find that the number of bound states
is controlled by the position of the parameter $\nu$ with respect to a set of 
values $\nu_{min}$ at which a bound state gets out of the well,
i.e. disappears. A careful study of this transcendental equation shows that
the parameter $\alpha$ of the bound state levels behaves as 
$\alpha_{max}/(1\,+\nu^{-1})$ when $\nu\to\infty$,
and as $\nu_{min}+\gamma (\nu-\nu_{min})$ when $\nu\to\nu_{min}$.
The three parameters ($\nu_{min}$, $\alpha_{max}$, $\gamma$) of these
asymptotic behaviours depend on the geometry of the confinement and 
the level under consideration.

$\bullet$~For quantum wells, the levels are characterized by one quantum number 
$n$, with $n=1,2,3,\cdots$, the corresponding parameters being
$\nu_{min}=(n-1)\pi/2$, \mbox{$\alpha_{max}=n\pi/2$} and $\gamma=1$.

$\bullet$~For cylindrical wires, the levels are characterized by two
quantum numbers $n$ and $\pm m$ with $m=0,1,2,\cdots$.
For the $(n,\pm m)$ levels, the parameter $\gamma$ is equal to $1/sup(m,1)$,
the parameter $\alpha_{max}$ is equal to
the $n^{th}$ zero $z^{(n)}_{m}$ of the Bessel function $J_{m}$
and the threshold $\nu_{min}$
for level disappearance is equal to $z^{(n)}_{m-1}$ for $m\neq 0$,
$z_{1}^{(n-1)}$ for $(m=0,n\neq 1)$ and $0$ for the ground state $(m=0, n=1)$.

$\bullet$~For spherical dots, the levels are characterized by three quantum numbers
$(n,l,\pm m)$ with $l=0,1,2,\cdots$; they are degenerate with respect to $m$.
For $l=0$, the parameters are \mbox{$\nu_{min}=(n-1/2)\pi$},
$\alpha_{max}=n\pi$ and $\gamma=1$, while for $l\neq 0$, they are 
$\nu_{min}=z^{(n)}_{l-1/2}$, $\alpha_{max}=z^{(n)}_{l+1/2}$ and $\gamma=1/(l+1/2)$. 

\begin{figure}[!h]
\epsfxsize=7cm
\epsfysize=7cm
\epsffile{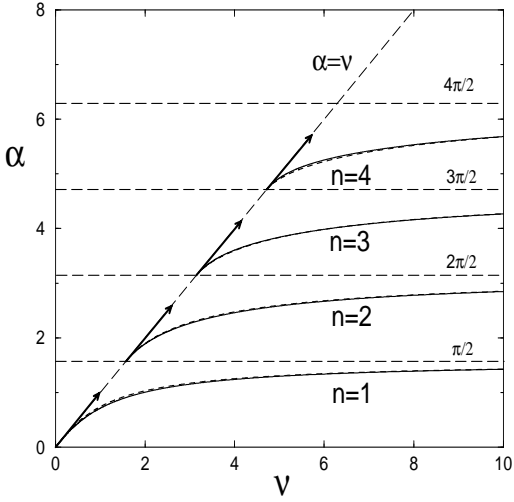}
\vglue 0.2cm
\caption{
\label{fig1}
The energy parameter $\alpha$ as defined in Eq.(\ref{eq2}) for the
$n^{th}$ bound states of a quantum well as obtained from the exact expression
of Eq.(\ref{eq9bis}) (solid line) and from the analytical expression
Eq.(\ref{eq4}) ( dashed line). The confinement parameter $\nu$
depends on the barrier height $V$ and {\it half width} $R$ through Eq.(\ref{eq1}
).
}
\end{figure}

\begin{figure}[!h]
\epsfxsize=7cm
\epsfysize=7cm
\epsffile{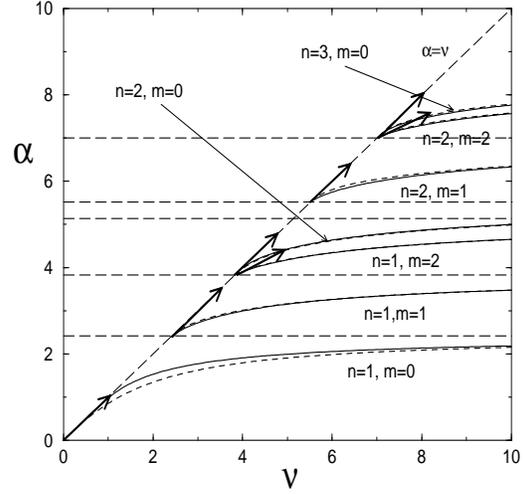}
\vglue 0.2cm
\caption{
\label{fig2}
The energy parameters $\alpha$ for the $n^{th}$ level of the $\pm m$ states
of the cylindrical wire confinement, as given by the numerical resolution
of Eq.(\ref{eq7},\ref{eq11}) (solid line)
and by the approximate analytical expressions
deduced from Eq.(\ref{eq3}) (dashed line).
The horizontal dashed lines correspond to $z^{(1)}_{0}\simeq 2.40$,
$z^{(1)}_{1}\simeq 3.83$, $z^{(1)}_{2}\simeq 5.13$, $z^{(2)}_{0}\simeq 5.52$,
$z^{(2)}_{1}\simeq 7.01$ with
$z^{(n)}_{m}$ being the $n^{th}$ zero of the Bessel function $J_m$.
 The confinement parameter $\nu$ depends on the barrier height and the cylinder
 radius $R$ through Eq.(\ref{eq1}). Note that two curves
$\alpha^{(n)}_{m=2}$ and $\alpha^{(n+1)}_{m=0}$ start each at
$\nu\simeq z^{(n)}_{1}$ (with different slopes).
}
\end{figure}

\begin{figure}[!h]
\epsfxsize=6.5cm
\epsfysize=6.5cm
\epsffile{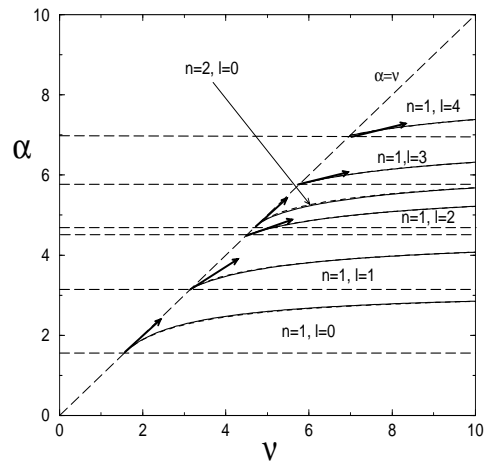}
\caption{                                      
\label{fig3}
The energy parameters $\alpha$ for the $n^{th}$ level of the $(l,m)$ states
of a spherical dot as given by the numerical resolution  of
Eq.(\ref{eq7},\ref{eq14})
or (\ref{eq16}) (solid line) or by the analytical expression deduced from Eq.(\r
ef{eq3})
or given in Eq.(\ref{eq5}).
The horizontal dashed lines correspond to $\pi/2$, $\pi$,
$z^{(1)}_{3/2}\simeq 4.49$, $3\pi/2$, $z^{(1)}_{5/2}\simeq 5.76$,
$z^{(1)}_{7/2}\simeq 6.99$ with
$z^{n}_{l+1/2}$ being the $n^{th}$ zero of
the Bessel function $J_{l+1/2}$. The confinement parameter $\nu$ depends on the
barrier height and
dot radius through Eq.(\ref{eq1}). Note that for this 3D confinement,
the confinement parameter has to be larger than $\pi/2$ for a bound state
to exist.
}
\end{figure}

From the numerical resolution of the transcendental equations
verified by the energy levels of these three confined geometries, we have checked
that the energy parameter $\alpha$ is amazingly well reproduced,
for {\it all} levels and {\it all} confinement $\nu$, by the same function
\bea\label{eq3}
\chi(\nu)&=&
\frac{\alpha_{max}\nu}
{
  \nu +1 +{\large\frac{(\alpha_{max}-\nu_{min}-1)^2}
 {(\alpha_{max}-\nu_{min}-1)+(\nu-\nu_{min})(1+(\gamma-1)\alpha_{max}/\nu_{min})
  }}
}\nonumber\\
&&
\eea
which is constructed to give the two first terms of the $\alpha$ behavior for
both $\nu\to\infty$ and $\nu\to\nu_{min}$ 
(see Fig.\ \ref{fig1},\ref{fig2},\ref{fig3}). The solid lines which correspond
to the numerical solutions of the energy transcendental equations
are hard to distinguish from the dashed lines which correspond to 
Eq.(\ref{eq3}): the discrepancy is indeed extremely small.

In the particular case of the $n^{th}$ level of a quantum well (which exits for 
$\nu>(n-1)\pi/2$ only), this gives

\bq\label{eq4}
\alpha^{(n)}(\nu)\simeq\frac{n\pi}{2}\frac{\displaystyle\nu}
{\displaystyle
  \nu +1 +\frac{\displaystyle(\pi/2-1)^2}
  {(\pi/2-1)+(\nu-(n-1)\pi/2)}
}
\ee
while for the $n^{th}$ level of the $l=0$ states of a spherical
dot (which exists for $\nu>(n-1/2)\pi$ only) this gives

\bq\label{eq5}
\alpha^{(n)}_{l=0}(\nu)\simeq n\pi\frac{\displaystyle\nu}
{\displaystyle
  \nu +1 +\frac{\displaystyle(\pi/2-1)^2}
  {(\pi/2-1)+(\nu-(n-1/2)\pi)}
}
\ee

\textbf{\underline{2. Derivation:}}

Let us now outline how we have derived the above results. The Schr\"odinger 
equation of a particle of mass $m$ in the confined
geometries considered here reads, in terms of the reduced variable
$\xi=z/R$ for quantum wells, $\vec{\xi}=(|\vec{\rho}|/R,\varphi)$
for cylindrical wires and $\vec{\xi}=(|\vec{r}|/R,\theta,\varphi)$ for
spherical dot, as
\bq\label{eq6}
\left[\Delta_{\vec{\xi}} + v(\xi)\right]\psi(\vec{\xi})=0
\ee 
with $v(\xi)=\alpha^2$ for $|\vec{\xi}|<1$ and 
$v(\xi)=-\beta^2$ for $|\vec{\xi}|>1$.

The bound states of these confined geometries correspond to $(\alpha,\beta)$
real and positive \cite{abpos}, $\alpha$ and $\beta$ being linked by:

\bq \label{eq7}
\alpha^2 +\beta^2 = \nu^2
\ee 
 due to Eq.(\ref{eq2}).

\vspace{.5cm}

\underline{a) 1D case: quantum well}

In the 1D case, the solution of Eq.(\ref{eq6}) is elementary:
the wavefunction which cancels at $\pm \infty$ and has continuous
derivatives at $\pm 1$ can be written as:
\bea\label{eq8}
\psi(-1<\xi<1)&=&a\left[e^{2i\alpha}\,e^{i\alpha\xi}+
e^{2i\arctan \frac{\beta}{\alpha}}\,e^{-i\alpha\xi}\right]\nonumber\\
\psi(\xi<-1)&=&\psi(-1)e^{\beta(\xi+1)}\\
\psi(\xi>1)&=&\psi(1)e^{-\beta(\xi-1)}\nonumber
\eea
with $\alpha$ and $\beta$ linked by Eq.(\ref{eq7}) and by
\bq\label{eq9}
\exp 4\,i\,\alpha = \exp(4\,i\,\arctan \beta/\alpha)
\ee
Eq.(\ref{eq9}) is not one of the standard forms of the quantum
well transcendental equation found in usual textbooks \cite{bastard,rosencher}. It is however the most
convenient one to extract the asymptotic behaviour of $\alpha$ as we now 
show.

(\textit{i})~ When $\nu\to\infty$, the finite values of $\alpha$ correspond
to $\beta\simeq\nu$ infinite. For such $\nu$, $\beta/\alpha$ is large
so that $\arctan \beta/\alpha\simeq \pi/2-\alpha/\beta$.
Eq.(\ref{eq9}) then leads to $4\,\alpha\simeq 4(\pi/2-\alpha/\nu)+2(n-1)\pi$
i.e. $\alpha\simeq \frac{n\pi}{2}/(1+\nu^{-1})$ with $n=1,2,3,\cdots$.

(\textit{ii})~The solutions for the bound state disappearance, i.e for an energy level
close to the top of the well, correspond to $\alpha\simeq\nu$ and 
$\beta\simeq 0$. $\beta/\nu$ is then small so that  Eq.(\ref{eq9}) gives
$\beta\simeq\nu(\nu-(n-1)\pi/2)$ i.e. 
$\alpha\simeq\nu+{\mathcal O}((\nu-(n-1)\pi/2)^2)$ due to Eq.(\ref{eq7}).
This gives the values of $\alpha_{max}$, $\nu_{min}$ and $\gamma$ listed above.

In order to check the validity of the approximate 
$\alpha^{(n)}(\nu)$ given in Eq.(\ref{eq4}), it is not
necessary to solve Eq.(\ref{eq7},\ref{eq9}) numerically:
indeed, while these equations do not give $\alpha$ in terms of $\nu$,
they do give $\nu$ in terms of $\alpha$ as
\bq\label{eq9bis}
\nu=\alpha/\cos(\alpha-(n-1)\pi/2)
\ee 

The corresponding curves as well as the various $\alpha^{(n)}(\nu)$ are shown
in Fig.\ \ref{fig1}: The fit is impressive; the two sets of curves
are hard to distinguish, the largest discrepancy is smaller than $4\%$. 

\vspace{.5cm}

\underline{b) 2D case: cylindrical wire}

In the 2D case, we have $\Delta_{\vec{\xi}}=
\frac{1}{\xi}\frac{\partial}{\partial\xi}\xi\frac{\partial}{\partial\xi}
+\frac{1}{\xi^2}\frac{\partial^2}{\partial\varphi^2}$,
so that the wave functions write 
$\psi(\vec{\xi})=f_{m}(\xi)\,e^{\pm i\,m\varphi}$
with $m=0,1,2,\cdots$. The differential equation obtained for $f_{m}$
is just the one of Bessel functions $(J_{m},Y_{m})$ for $\xi$ inside
the well and $(I_{m},K_{m})$ for $\xi$ outside.
The solutions which are finite for $\xi=0$, tend to $0$ for 
$\xi\to\infty$ and have a continuous derivative for $\xi=1$ read:
\bea\label{eq10}
f_{m}(\xi<1)&=&A\,J_{m}(\alpha\,\xi)\nonumber\\
f_{m}(\xi>1)&=&A\,\frac{J_{m}(\alpha)}{K_{m}(\beta)}K_{m}(\beta\,\xi)
\eea
with $\alpha$ and $\beta$ linked by Eq.(\ref{eq7}) and by
\bq\label{eq11}
\frac{\alpha_{m}\,J_{m-1}(\alpha_{m})}{J_{m}(\alpha_{m})}=
-\frac{\beta_{m}\,K_{m-1}(\beta_{m})}{K_{m}(\beta_{m})}
\ee
 In order to get the above expression,
we have used $J_{-1}=-J_{1}$ and $K_{-1}=K_{1}$ and \cite{absteg}:

\bea\label{eq12}
\textrm{(a)}\, J^{\prime}_{\nu}&=&\quad J_{\nu-1}-\nu/z\,J_{\nu}\nonumber\\
\qquad\textrm{(b)}\, K^{\prime}_{\nu}&=&-K_{\nu-1}-\nu/z\,K_{\nu}
\eea

(\textit{i})~When $\nu\to\infty$, the finite values
of $\alpha_{m}$ correspond to $\beta\simeq\nu$ infinite. The $K$ ratio being 
$1$ in this limit, we find 
$J_{m}(\alpha_{m})/\alpha_{m}\,J_{m-1}(\alpha_m)\simeq -1/\nu\simeq 0$.
The values of $\alpha_m$ are thus close to the zeros of $J_m$.
For these $\alpha_{m}$, we find using Eq.(\ref{eq12}.a) that the inverse
of the LHS of Eq.(\ref{eq11}) is close to
$(\alpha_{m}-z^{(n)}_{m})/z^{(n)}_{m}$ so that 
$\alpha_m\simeq z^{(n)}_{m}/(1+\nu^{-1})$. 

(\textit{ii})~The solutions close to the top of the well
correspond to $\beta_{m}\simeq 0$ and $\alpha_{m}\simeq \nu$. For these
$\beta_{m}\simeq 0$ the RHS of Eq.(\ref{eq11}) is close to
$1/\ln \beta$ for $m=0$, $\beta^2 \ln \beta$ for $m=1$
and $-\beta^2 /2(m-1)$ for $m\ge 2$.
Being close to zero in all cases, we conclude that $\alpha_m$ must tend to 
the zeros of $J_{m-1}$. In the particular case of $m=0$, we find that
in addition to the zeros of $J_{-1}$, i.e. the zeros of $J_{1}$, $\alpha_{0}$
can also tend to $0$ for $\nu\to 0$. Since in this $\beta\sim\nu$ limit, 
$\alpha\sim\nu$ while the LHS of Eq.(\ref{eq11}) is close to
$-z^{(n)}_{m-1}(\alpha-z^{(n)}_{m-1})$ due to Eq.(\ref{eq12}a),
we get for $m\ge 1$, $\alpha_{m}\simeq z^{(n)}_{m-1}+(\nu-z^{(n)}_{m-1})/m$
while for $m=0$, we find $\alpha_{0}\simeq \nu_{min}+(\nu-\nu_{min})$
with $\nu_{min}=0$ or $z^{(n)}_1$.

We thus recover the values of $(\nu_{min},\alpha_{max},\gamma)$ listed above.
Fig.\ref{fig2} shows the results of the numerical resolution of
Eqs.(\ref{eq7},\ref{eq11}) as well as the value of
$\alpha^{(n)}_{m}$ deduced from the corresponding 
$(\nu_{min},\alpha_{max},\gamma)$ inserted in Eq.(\ref{eq3}).
The fit is excellent for all confinements, the two curves being hard to
distinguish, with an exception for the ground state  at intermediate $\nu$:
the approximate $\alpha^{(n=1)}_{m=0}$ is slightly below the numerical
$\alpha$, the largest discrepancy being $14\%$.

\underline{c) 3D case : spherical dot}

In the 3D case, we have 
$\Delta_{\vec{\xi}}=
\frac{1}{\xi^2}\frac{\partial}{\partial\xi}\xi^2\frac{\partial}{\partial\xi}
-\vec{L}^2/\xi^2$, so that the wave functions write
$\psi(\vec{\xi})=f_{l}(\xi)Y_{l,m}(\theta,\varphi)$,
with $l=0,1,2,\cdots$ and $-l\leq m\leq l$. 
By setting $f_{l}=g_{l}/\sqrt{\xi}$, we find that the differential equation
for $g_l$ is just the one of Bessel functions $(J_{l+1/2},Y_{l+1/2})$ for 
$\xi$ inside the well, and $(I_{l+1/2},K_{l+1/2})$ for $\xi$ outside.
The solution which are finite for $\xi=0$, tend to $0$ for $\xi\to\infty$
and have a continuous derivative for $\xi=1$ read:
\bea\label{eq13}
f_{l}(\xi<1)&=&\frac{A}{\sqrt{\xi}}\,J_{l+1/2}(\alpha\,\xi)\nonumber\\
f_{l}(\xi>1)&=&\frac{A}{\sqrt{\xi}}\frac{J_{l+1/2}(\alpha)}{K_{l+1/2}(\beta)}
\,K_{l+1/2}(\beta\,\xi)
\eea
with $\alpha$ and $\beta$ linked by Eq.(\ref{eq7}) and by
\bq\label{eq14}
\frac{\alpha_{l}\,J_{l-1/2}(\alpha_{l})}{J_{l+1/2}(\alpha_{l})}=
-\frac{\beta_{l}\,K_{l-1/2}(\beta_{l})}{K_{l+1/2}(\beta_{l})}
\ee
for $l\ge 1$ (due again to Eq.(\ref{eq12})). For the $l=0$ states however, it is simpler to use
the explicit expression $J_{1/2}(x)=(\frac{2}{\pi\,x})^{1/2} \sin x$ and 
\mbox{$K_{1/2}(x)=(\frac{\pi}{2\,x})^{1/2}\,e^{-x}$ } in Eq.(\ref{eq13}). 
The continuity of $f^{\prime}_{0}$ at $\xi=1$ then leads to
\bq\label{eq15}
\alpha_{0}/\tan\alpha_{0}=-\beta_{0}
\ee
so that due to Eq.(\ref{eq7}), the transcendental equation for the $l=0$
levels of a spherical dot is simply
\bq\label{eq16}
\nu=\alpha_{0}/|\sin \alpha_{0}|
\ee
with $(n-1/2)\pi<\alpha_{0}<n\pi$ in order to have $\beta_{0}>0$.\\
Let us start with these $l=0$ states.

(\textit{i})~For $\nu\to\infty$, Eq.(\ref{eq16}) leads to 
$|\sin \alpha_0 |\to 0$ so that $\alpha_{0}\simeq n\pi$. The expansion
of the RHS of Eq.(\ref{eq16}) close to these maximum values of $\alpha$ gives
\mbox{$\alpha_{0}\simeq n\pi/(1+\nu^{-1})$.} 

(\textit{ii})~For $\beta_{0}\simeq 0$, Eq.(\ref{eq15}) leads to 
$\cos\alpha_{0}\simeq 0$; so that $\alpha_{0}\simeq (n-1/2)\pi$ only
in order to have $\beta_{0}>0$. The expansion of the LHS of Eq.(\ref{eq15})
close to these values gives 
$\alpha_{0}\simeq \nu+{\mathcal O}((\nu-(n-1/2)\pi)^2)$.

We now turn to the $l\ge 1$ states.

(\textit{i})~For $\nu\to\infty$ and $\alpha_l$ finite, i.e.
$\beta_{l}\simeq\nu$ infinite, the ratio of $K$ is equivalent to $1$, so that
$J_{l+1/2}(\alpha_{l})/\alpha_{l}J_{l-1/2}(\alpha_{l})
\simeq -1/\beta_{l}\sim 0$. $\alpha_{l}$ is thus close to a zero of 
$J_{l+1/2}$. Using Eq.(\ref{eq12}a), the expansion of the inverse of the
LHS of (\ref{eq14}) leads to $(\alpha_{l}-z^{(n)}_{l+1/2})/z^{(n)}_{l+1/2}$
so that $\alpha_{l\ne 0}\simeq z^{(n)}_{l+1/2}/(1+\nu^{-1})$. 

(\textit{ii})~For $\beta_{l}\to 0$, and $\alpha_{l}\simeq\nu$ finite,
the RHS of Eq.(\ref{eq14}) goes to zero as $-\beta_{l}^2/(2l-1)$ so that 
$\alpha_{l}$ is close to a zero of $J_{l-1/2}$. Close to these values, the LHS 
of Eq.(\ref{eq14}) is equivalent to 
$z^{(n)}_{l-1/2}(\alpha_{l}-z^{(n)}_{l-1/2})$ due to Eq.(\ref{eq12}a)
so that in this limit we have\mbox{
$\alpha_{l\ne 0}\simeq z^{(n)}_{l-1/2} +(\nu-z^{(n)}_{l-1/2})/(l+1/2)$.}

This just corresponds to the values of $(\nu_{min},\alpha_{max},\gamma)$
listed above. Fig.\ref{fig3} shows the results of the numerical resolution
of Eq.(\ref{eq7},\ref{eq14}) or (\ref{eq16}) as well as the values of $\alpha$
deduced from Eq.(\ref{eq3}) or given in equation (\ref{eq5}).
Here again, the fit is increadibly good for all confinement 
and all energy levels.

\vspace{.2cm}

\underline{d) One specific example:}

The above results are of course given in reduced units in order to be universal.
Let us apply them to one specific case: $GaAs/Ga_{x}Al_{1-x}As$.
For usual $x$'s, the electron mass is $m\simeq 0.07$ and the energy barrier
for electrons is of the order of $V\simeq 300\, meV$.
For quantum well, wire or dot of radius $R\simeq 40$\AA, Eq.(\ref{eq1}) gives
the confinement parameter $\nu\simeq 3.0$. \mbox{Figure(\ref{fig1},\ref{fig3})}
show that two bound states exist for quantum well or wire, but one only
for quantum dot. Moreover the energies of these levels are significantlty lower
than the energies of the corresponding states for infinite barrier:
in the case of a quantum well of half width $40$\AA, Eq.(\ref{eq4}) gives
$\alpha\simeq 1.1$ and $2.24$ instead of $\pi/2$ and $\pi$ which corresponds 
to $45$ and $170$ meV instead of $85$ and $340$ meV.

\textbf{\underline{In conclusion}}, we have given analytical expressions for the bound states
energies of quantum wells, cylindrical wires, and spherical dots,
which reproduce impressively well their numerical values,
for {\it any} level, barrier height and confinement size, without adjustable
parameters.
From them, we can easily obtain the exact wave functions of
these bound states by inserting the corresponding values
of $(\alpha,\beta)$ into Eq.(\ref{eq8},\ref{eq10},\ref{eq13}).

* Laboratoire associ\'e au Centre National
de la Recherche Scientifique et aux Universit\'es Paris 6 et \mbox{Paris 7.}


\begin{references}
\bibitem{abpos} The condition $(\alpha,\beta)$ positive is useful to identify
the wavefunction which fulfills the boundary condition for $\xi\to \infty$.
\end{references}
\end{document}